\documentclass[final,3p]{elsarticle}

\usepackage{amssymb}
\usepackage{amsmath}
\usepackage{amsthm}
\usepackage{comment}
\usepackage{xcolor} 
\usepackage{graphicx}
\usepackage{epstopdf}
\usepackage{caption}
\usepackage{subcaption}
\usepackage{wrapfig}
\usepackage{wrapfig,lipsum,booktabs}

\journal{Mechanics Research Communications}

\begin{document}

\begin{frontmatter}



\title{Capillary Rise in Pipes with Variable Cross Section\tnoteref{ack}}
\tnotetext[ack]{The authors gratefully acknowledge the financial support of the Ministry of Science, Technological Development and Innovation of the Republic of Serbia, Grant No. 451-03-137/2025-03/200125 and 451-03-136/2025-03/200125 (S. Simi\'{c}) and through the Mathematical Institute of the Serbian Academy of Sciences and Arts (I. Rapaji\'{c}).}

\author[a,b]{Isidora Rapaji\'{c}} 
\ead{isidora.rapajic@turing.mi.sanu.ac.rs}
\affiliation[a]{organization={Mathematical Institute of the Serbian Academy of Sciences and Arts},
            addressline={Kneza Mihaila 36},
            city={11001 Belgrade},
            country={Serbia}}

\author[b]{Srboljub Simi\'{c}\corref{cor1}} 
\ead{ssimic@uns.ac.rs} 
\affiliation[b]{organization={Department of Mathematics and Informatics, Faculty of Sciences},
            addressline={Trg Dositeja Obradovica 4},
            city={21000 Novi Sad},
            country={Serbia}}

\cortext[cor1]{Corresponding author}


\begin{abstract}
This work proposes an ODE model for a capillary rise in pipes with variable cross section, and compares it to the lubrication theory model. Two key assumptions are made: (1) radius of the pipe varies with axial coordinate, and (2) pipe's convergence angle is small. The model reduction process involves the identification of critical parameters and simplifies the governing equations by neglecting higher-order terms. Under appropriate scaling, it is shown that generalized Washburn's equation for capillary rise in pipes with variable cross section reduces to the lubrication theory model \cite{Figliuzzi, Vella}.
\end{abstract}




\end{frontmatter}





\section{Introduction}


Capillary motion of the fluid through a narrow pipe is an ubiquitous phenomenon whose importance spans from biology (e.g. water uptake in plants) to industry. It also serves as a model for capillary imbibition through the porous medium. The first results relating the penetration length $l$ to time $t$ were given by Bell and Cameron in 1905 \cite{Bell-1906}, Lucas in 1918 \cite{lucas1918ueber}, and Washburn in 1921 \cite{Washburn}, giving
\begin{equation} \label{Intro:WashLaw}
	l(t) = K t^{1/2},
\end{equation}
where $K$ is a constant. This equation is a consequence of the balance between surface tension and viscous friction. 

It is obvious that \eqref{Intro:WashLaw} does not capture other effects that occur in capillary flow, such as inertia, gravity, or the dynamic contact angle of the free surface. One of the simplest, but complete models that describes the capillary rise of a fluid through a pipe with circular cross section of constant radius $R_{0}$ is known as Washburn's (or Lucas-Washburn's) equation. It was proposed by Bosanquet in 1923 \cite{Bosanquet}, using Newton's second law applied to a control volume. It reads: 
\begin{equation} \label{Intro:Washburn}
	\rho \frac{\mathrm{d}}{\mathrm{d}t} \left[ h(t) \frac{\mathrm{d}h(t)}{\mathrm{d}t} \right] 
	+ \frac{8 \mu}{R_{0}^{2}} h(t) \frac{\mathrm{d}h(t)}{\mathrm{d}t} 
	+ \rho g h(t) 
	= \frac{2 \gamma \cos \theta}{R_{0}}, 
\end{equation}
where $h(t)$ is the height of a fluid column (position of the meniscus) at time $t$. The fluid is assumed to be Newtonian, with mass density $\rho$, viscosity $\mu$, surface tension $\gamma$, and contact angle $\theta$ between the fluid free surface and wall of the pipe ($g$ denotes gravitational acceleration). All physical properties of the fluid are assumed constant. 

Washburn's equation \eqref{Intro:Washburn} is one of the rare models in which the motion of a viscous fluid, described by the Navier-Stokes equations, is reduced to a single ordinary differential equation (ODE). Simple inspection reveals that the Washburn relation \eqref{Intro:WashLaw} is a solution of \eqref{Intro:Washburn} when inertia and gravity are neglected. A detailed mathematical analysis of the existence, uniqueness, and behavior of the solution of \eqref{Intro:Washburn} is provided in \cite{Switala}, while comparisons to experimental data can be found in \cite{Quere, Bothe}. For a pipe with constant radius, the equation can be derived in polar-cylindrical coordinates from the Navier-Stokes equations \cite{Kornev}, with a similar procedure applied to the case of capillary flow between two parallel plates \cite{Grunding}. A recent study \cite{ruiz2022long} focused on the transition between different dynamical regimes in Washburn's equation with dynamic contact angle. 

A model that describes capillary flow through the pipe with variable circular cross section $R(z)$ has been derived in \cite{Figliuzzi,Vella}, and bears the following form: 
\begin{equation} 
	\begin{split}
		\left[ 8 \mu R(h(t))^{3} \int_{0}^{h(t)} R(z)^{-4} \mathrm{d}z \right]
		\frac{\mathrm{d}h(t)}{\mathrm{d}t} & + \rho g h(t) R(h(t)) \\
		& = 2 \gamma \operatorname{cos} \left( \theta +  \mathrm{arctan} \frac{\mathrm{d}R(h(t))}{\mathrm{d}h} \right).
	\end{split}
	\label{Intro:Lubrication}
\end{equation} 
It was derived using the `lubrication theory' approximation, which inherits the following assumptions: (i) cross-sectional representative length $R_{0}$ is much smaller than the length of the pipe $L$, $R_{0} \ll L$, which implies the long and thin flow domain; (ii) cross section varies slowly along the axial coordinate $z$, and (iii) conditions of the flow are such that inertial terms may be neglected. Although the variation of the cross section along the pipe axis is taken into account, equation \eqref{Intro:Lubrication} does not generalize the Washburn model \eqref{Intro:Washburn} completely. It does not take into account inertial terms, which imposes limitations to the applicability of the model. 

\begin{figure}[ht]
	\centering
	\includegraphics[width=0.5\textwidth]{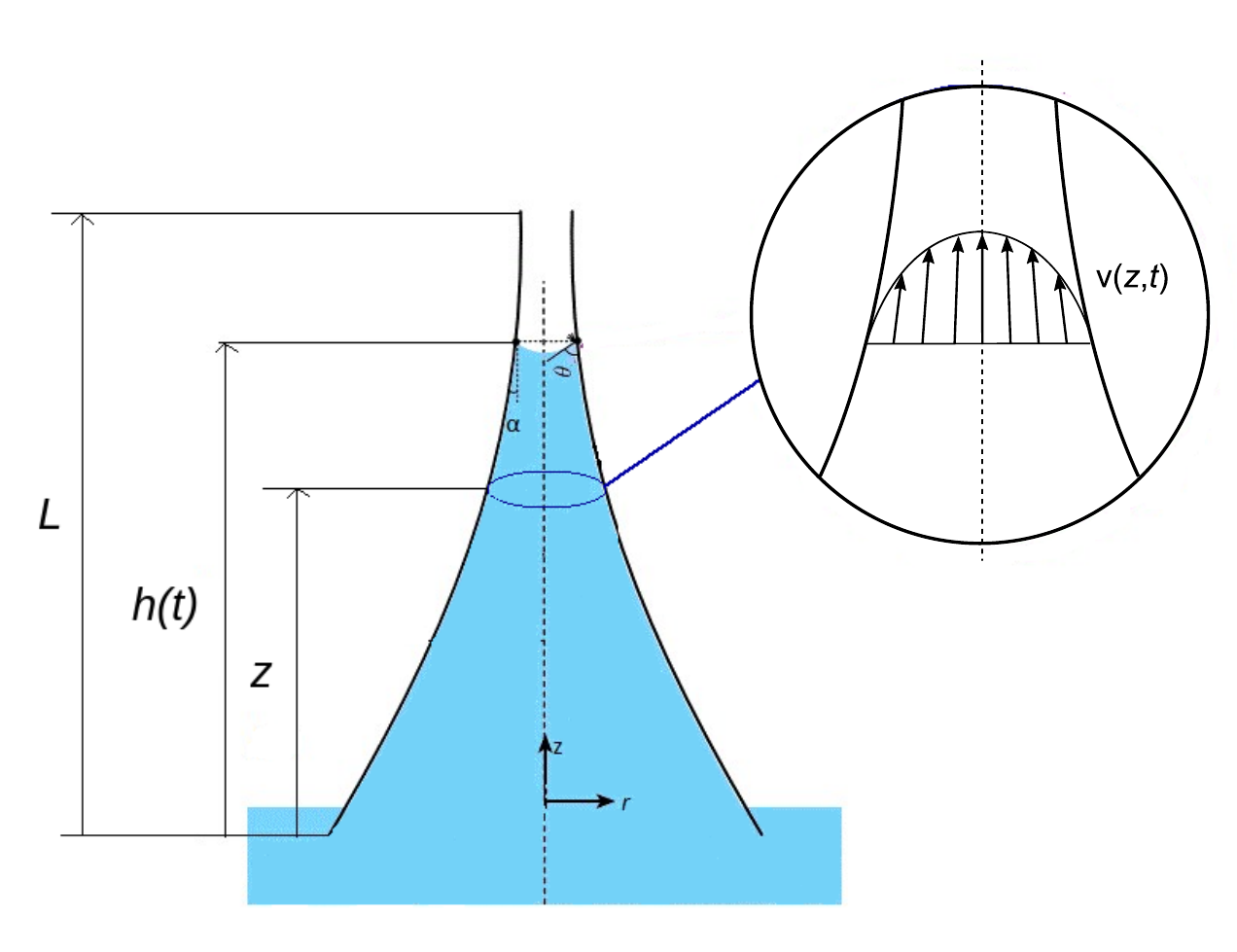}
	\caption{Poiseuille flow in a capillary pipe of variable radius, with meniscus height $h(t)$ at time $t$.}
	\label{fig::Pipe}
\end{figure}

In view of the shortcomings of the Washburn equation (constant cross section) and the lubrication theory model (absence of inertial terms), this study aims to generalize the Washburn equation \eqref{Intro:Washburn} to the case of variable circular cross section. To make the model as general as possible, our approach will be based upon basic equations---mass and momentum balance laws in local and global form. This will lead to retaining the inertial term. Since the target model is supposed to be an ordinary differential equation, assumption of a slowly varying cross section along the long and thin domain will be kept. Moreover, careful use of the global form of balance laws, and systematic application of asymptotic analysis, will facilitate consistent generalization of both aforementioned models. Although the resulting model will yield similar results to the lubrication theory model \eqref{Intro:Lubrication} in the case of negligible inertia effects, it will also capture the oscillatory behaviour which cannot be recovered in the model without inertial terms. To that end, we claim that our model is more complete than the ones used so far. Consequently, it is more appropriate for further analysis of existence and uniqueness of solutions, and their stability analysis. 

The rest of the paper is organized as follows. In Section \ref{sec:Assumptions-Local}, the assumptions will be listed along with the local analysis of the Navier-Stokes equations. In Section \ref{sec:Global}, a global analysis of the momentum balance law will be given, leading to the generalized Washburn equation \eqref{eq::GeneralisedWashburn}. In addition, a comparison of the generalized Washburn equation \eqref{eq::GeneralisedWashburn} with the lubrication theory model \eqref{Intro:Lubrication} will be given, based on a formal asymptotic analysis. Finally, in Section \ref{sec:ComparisonNum}, numerical solutions of different models will be compared. Conclusions and perspective for further study will close the paper. Technical details of the computations are provided in the Appendix. 

\section{Assumptions and local analysis} \label{sec:Assumptions-Local} 


Derivation of the Washburn equation is usually relied on certain simplified approaches whose basis is the momentum equation, i.e. Newton's Second Law. There are only a few examples in which the analysis possesses a higher level of rigor \cite{Kornev, Grunding}. Since simplified procedures cannot be generalized in a straightforward way, our analysis will be based upon the basic equations of fluid mechanics---mass and momentum balance laws. To that end, we need a precise list of assumptions which will admit simplification of the balance laws in local form, i.e. Navier-Stokes equations. 

\subsection{Assumptions} 

Assumptions in this model are of geometrical, kinematical, and physical kind. 

\begin{enumerate}
	\item \emph{Geometry of the pipe.} It is assumed that the pipe has a variable circular cross section of radius $R(z)$, with $R_{0} = R(0)$. Main assumptions regarding the cross section are the following: (i) cross-sectional radius varies slowly with axis coordinate $z$, and (ii) cross-sectional radius decreases with $z$. They can be expressed as follows:
	\begin{equation} \label{eq::slowlyVarradius}
		|R'(z)| = |\tan \alpha| \ll 1, \quad 
		R'(z) < 0.
	\end{equation}
	By assumption (ii) the analysis is restricted to converging pipes. Other possible shapes of the pipes, e.g. sinusoidal ones, will not be the subject of the present study. 
	
	\item \emph{Velocity field.} It is assumed that the fluid flow is axially symmetric with respect to $z-$axis. This implies that the velocity field in polar-cylindrical coordinates is independent of the angular coordinate $\varphi$, $\mathbf{v} = \mathbf{v}(r,z,t)$, and has the form: 
	\begin{equation} \label{eq::VelocityField}
		\mathbf{v} = v_r \mathbf{e}_r + v_z \mathbf{e}_z, \quad 
		v_\varphi = 0.
	\end{equation}
	It is also assumed that the axial component of the velocity is described by the modulated Poiseuille flow profile, with no-slip at the wall (see Figure \ref{fig::Pipe}):
	\begin{equation} \label{eq::PoiseuilleFlow}
		v_{z}(r,z,t) = v(z,t) \left( 1 - \frac{r^{2}}{R(z)^{2}} \right). 
	\end{equation}
	This assumption relies on a geometrical assumption \eqref{eq::slowlyVarradius}. Note that the standard parabolic profile in \eqref{eq::PoiseuilleFlow} is $z-$dependent due to variable radius. It is modulated by $v(z,t)$, which is assumed to be bounded for $t \in [0,\infty)$. Such a modulation is necessary to preserve mass conservation. Furthermore, it will be assumed that modulation varies slowly along the axis:
	\begin{equation} \label{eq::slowlyVarModulation}
		\left| \frac{\partial v(z,t)}{\partial z} \right| \ll 1, 
	\end{equation}
	again in relation to \eqref{eq::slowlyVarradius}. 
	
	\textit{Remark.} Note that in \cite{Vella, Bothe}, the velocity field profile was derived from the momentum balance law with the inertial term neglected. In those works, the pressure gradient appears explicitly in the velocity field, as it serves as the driving force of the flow. In contrast, our analysis considers the full momentum balance, including the inertial term, and the pressure gradient is incorporated into the equations through the contact force.
	
	\item \emph{Physical properties of the fluid.} Fluid is assumed to be Newtonian, i.e. incompressible and viscous. All physical properties, $\rho$, $\mu$, $\gamma$, and $\theta$, are assumed constant. 
\end{enumerate}
Although \eqref{eq::slowlyVarradius} is merely a geometrical assumption, it has substantial influence on the model. In the sequel, we shall use the following order of magnitude estimates:
\begin{equation} \label{eq::OrderOfMagnitude}
	|R'(z)| = \mathcal{O}(\alpha), \quad 
	\left| \frac{\partial v(z,t)}{\partial z} \right| = \mathcal{O}(\alpha), \quad 
	\mathcal{O}(\alpha) \ll 1. 
\end{equation}
It will be assumed that terms of order $\mathcal{O}(\alpha)$ may be neglected in the analysis.  

\subsection{Local analysis} 

Local analysis is concerned with the consequences of assumptions on the mass and momentum balance laws in local form (i.e. Navier-Stokes equations). 

\subsubsection{Mass balance}

Since the fluid is assumed to be incompressible, mass balance law is reduced to continuity equation, $\operatorname{div} \mathbf{v} = 0$, which in polar-cylindrical coordinates reads: 
\begin{equation}
	\frac{1}{r} \frac{\partial}{\partial r}(r v_r) 
	+ \frac{1}{r} \frac{\partial v_{\varphi}}{\partial \varphi} 
	+ \frac{\partial v_z}{\partial z} = 0.
	\label{eq::massBalanceCylindrical}
\end{equation}
Assumption \eqref{eq::PoiseuilleFlow}, along with order of magnitude estimate \eqref{eq::OrderOfMagnitude}, implies: 
\begin{equation*}
	\frac{\partial v_z}{\partial z} = 
	\underbrace{\frac{\partial v(z,t)}{\partial z}}_{\mathcal{O}(\alpha)} 
	\left( 1 - \frac{r^{2}}{R(z)^{2}} \right)
	+ 2 v(z,t) \frac{r^2}{R^3(z)} 
	\underbrace{R'(z)}_{\mathcal{O}(\alpha)} 
	= \mathcal{O}(\alpha), 
\end{equation*}
when $v(z,t)$ is bounded. Taking into account assumption \eqref{eq::VelocityField}, one obtains from \eqref{eq::massBalanceCylindrical}: 
\begin{equation} \label{eq::RadialVelocity}
	v_r = \frac{1}{r} \int_{0}^{r} - s \frac{\partial v_{z}(s,z,t)}{\partial z} \mathrm{d}s
	= \frac{1}{r} \int_{0}^{r} - 2 v(z,t) \frac{s^3}{R^3(z)}R'(z) \mathrm{d}s 
	= \mathcal{O}(\alpha) 
\end{equation}
It follows that velocity field can be approximated as:
\begin{equation} \label{eq::VelocityFieldApprox}
	\mathbf{v} = v_z(r,z,t) \mathbf{e}_{z} + \mathcal{O}(\alpha),
\end{equation}
for $v_{z}(r,z,t)$ given by \eqref{eq::PoiseuilleFlow}. 

\subsubsection{Momentum balance} 

Momentum balance law for incompressible fluid in local form reads: 
\begin{equation} \label{eq::MomentumBalance}
	\frac{\partial \mathbf{v}}{\partial t} + (\mathbf{v} \cdot \nabla) \mathbf{v} 
	= - \frac{1}{\rho} \nabla p + \frac{\mu}{\rho} \nabla^{2} \mathbf{v} + \mathbf{g}. 
\end{equation}
For the sake of brevity we shall omit the component form of the momentum balance law in polar-cylindrical coordinates, which may be found in the literature \cite{acheson1990elementary}. Nevertheless, simple inspection of radial and angular component of \eqref{eq::MomentumBalance} reveals: 
\begin{equation*}
	\frac{\partial p}{\partial r} = \mathcal{O}(\alpha), \quad 
	\frac{\partial p}{\partial \varphi} = \mathcal{O}(\alpha). 
\end{equation*}
Assuming that pressure field is stationary (independent of $t$), one arrives to the following approximation: 
\begin{equation} \label{eq::PressureField}
	p = p(z) + \mathcal{O}(\alpha). 
\end{equation}
This will be the only consequence of the local momentum balance law to be used in the sequel. 

As a final remark, let us note that the velocity field \eqref{eq::VelocityFieldApprox} with the Poiseuille flow profile \eqref{eq::PoiseuilleFlow}, and the pressure field \eqref{eq::PressureField} represent only approximate solution of the local form of momentum balance law \eqref{eq::MomentumBalance}. Moreover, this solution still inherits field variables---velocity \eqref{eq::VelocityFieldApprox} depends on both time and space variables---and the model is still in the form of partial differential equation. To generalize the Washburn equation in the form of ODE, we have to pursue with a global analysis of mass and momentum balance. 


\section{Global analysis and the generalized Washburn equation} \label{sec:Global} 


Equation that determines dynamics of a fluid column is the momentum balance law \eqref{eq::MomentumBalance}. However, to reduce it to a single ODE, with time $t$ as an independent variable, field variables have to be averaged out in a certain sense. To reach this goal, the following strategy will be adopted: (i) the mean velocity of the circular cross section (perpendicular to the pipe axis) will be introduced; (ii) the momentum balance law in global form will be used for dynamical analysis. These two steps will eventually yield the generalized Washburn equation. 


\subsection{Global order of magnitude estimates} 

Global order of magnitude analysis is a necessary step in derivation of the generalized Washburn equation. Even more, it will be used to facilitate comparison of the generalized Washburn equation with the lubrication theory model. Therefore, subsequent analysis will rely on local order of magnitude estimates \eqref{eq::OrderOfMagnitude}.

First estimate is of a geometrical kind. It estimates the integral of a power law function of the cross section radius along the pipe axis. Simple integration by parts yields: 
\begin{align} \label{eq::EstimateRadius}
	\int_{0}^{h(t)} R(z)^{n} \, \mathrm{d}z 
	& = z R(z)^{n} \Big|_0^{h(t)} - \int_{0}^{h(t)} n z R(z)^{n-1} R'(z) \, \mathrm{d}z 
	\\ 
	& = h(t) R(h(t))^{n} + \mathcal{O}(\alpha). 
	\nonumber
\end{align} 
Other estimates are concerned with a velocity field, i.e. the modulating factor $v(z,t)$ along the pipe axis. In particular, we shall exploit the following estimates, again derived using integration by parts: 
\begin{align} \label{eq::EstimateVelocity}
	\int_{0}^{h(t)} v(z,t) \, \mathrm{d}z 
	& = z v(z,t) \Big|_{0}^{h(t)} 
	  - \int_{0}^{h(t)} z \frac{\partial v(z,t)}{\partial z} \, \mathrm{d}z 
	\\ 
	& = h(t) v(h(t),t) + \mathcal{O}(\alpha), 
	\nonumber
\end{align} 
\begin{align} \label{eq::EstimateFlux}
	\int_{0}^{h(t)} v(z,t) R(z)^{2} \, \mathrm{d}z 
	& = z v(z,t) R(z)^{2} \Big|_{0}^{h(t)} 
	  - \int_{0}^{h(t)} \left( z \frac{\partial v(z,t)}{\partial z} R(z)^{2} 
	  + 2 z v(z,t) R(z) R'(z) \right) \mathrm{d}z 
	\\ 
	& = h(t) v(h(t),t) R(h(t))^{2} + \mathcal{O}(\alpha), 
	\nonumber
\end{align} 
In \eqref{eq::EstimateRadius}-\eqref{eq::EstimateFlux} we used the local estimates \eqref{eq::OrderOfMagnitude}. Note that global velocity estimates (along the pipe axis $z$) lead to localization. However, they do not remove the velocity field from the problem. They simply evaluate it at the meniscus. We shall need one more step before reaching our goal---generalization of the Washburn equation. 

\subsection{Global mass balance and mean velocity of the meniscus} 

Global mass balance for incompressible fluids implies the continuity equation. In our problem, this reduces to the fact that volumetric flux $Q$ through any cross section perpendicular to the pipe axis is constant. This can be evaluated by integration of the velocity $v_{z}(r,z,t)$, given by \eqref{eq::PoiseuilleFlow}, over the cross section: 
\begin{align} \label{eq::ContinuityEq}
	Q & = \int_{0}^{R(z)} \int_{0}^{2\pi} v_{z}(r,z,t) r \mathrm{d}r \mathrm{d}\varphi 
	\nonumber \\ 
	& = \int_{0}^{R(z)} \int_{0}^{2\pi} v(z,t) \left( 1 - \frac{r^{2}}{R(z)^{2}} \right) 
	  r \mathrm{d}r \mathrm{d}\varphi 
	\\ 
	& = \frac{1}{2} v(z,t) R(z)^{2} \pi. 
	\nonumber
\end{align}
In the case of constant cross section \cite{rapajic2025134842}, modulating factor is independent of $z$ coordinate, $v = v(t)$. Here, its $z$-dependence compensates the change of a cross-sectional radius $R(z)$. 

Our aim is to generalize Washburn's equation as ordinary differential equation (if possible). Therefore, using the arguments applied in derivation of continuity equation \eqref{eq::ContinuityEq}, we compute the mean velocity of the meniscus as the average $z$-component of the velocity at $z = h(t)$: 
\begin{equation} \label{eq::AverageVelocity}
	\overline{v}_{h}(t) = \frac{1}{R(h(t))^{2} \pi} 
	  \int_{0}^{R(h(t))} \int_{0}^{2\pi} v_{z}(r,h(t),t) r \mathrm{d}r \mathrm{d}\varphi 
	  = \frac{1}{2} v(h(t),t).
\end{equation}
This relation will be of utmost importance in further analysis. 

\subsection{Global momentum balance} 

Momentum balance law in global form reads: 
\begin{equation} \label{eq::MomentumGlobal}
	\frac{\mathrm{d}}{\mathrm{d}t} \int_{V(t)} \rho \mathbf{v} \, \mathrm{d}V 
	= \int_{\partial V(t)} \mathbf{t}(\mathbf{n}) \, \mathrm{d}S 
	+ \int_{V(t)} \rho \mathbf{b} \, \mathrm{d}V, 
\end{equation}
where $\mathbf{t}(\mathbf{n})$ denotes traction at the point of the boundary with outer unit normal $\mathbf{n}$, and $\mathbf{b} = - g \mathbf{e}_{z}$ is the body force per unit mass. Region $V(t)$, to which the balance law \eqref{eq::MomentumGlobal} is applied, is defined as: 
\begin{equation} \label{eq::Region}
	V(t) = \{ (r,\varphi,z) | r \in [0,R(z)], \varphi \in [0,2 \pi), z \in [0, h(t)] \}, 
\end{equation}
where $h(t)$ is the height of the fluid column. Boundary $\partial V(t)$ is a union of three surfaces $\partial V_{i}(t)$, $i = 1,2,3$, with outer unit normals $\mathbf{n}_{i}$: 
\begin{equation*}
	\partial V(t) = \partial V_{1}(t) \cup \partial V_{2}(t) \cup \partial V_{3}(t),
\end{equation*} 
defined as follows: 
\begin{alignat}{2} \label{eq::Boundary}
	\partial V_{1}(t) & = \{ (r,\varphi,z) | r \in [0,R_{0}], \varphi \in [0,2 \pi), z = 0 \}, 
	& \quad \mathbf{n}_{1} & = - \mathbf{e}_{z}, 
	\nonumber \\ 
	\partial V_{2}(t) & = \{ (r,\varphi,z) | r = R(z), \varphi \in [0,2 \pi), z \in [0,h(t)] \},  
	& \quad \mathbf{n}_{2} & = \cos \alpha \mathbf{e}_{r} + \sin \alpha \mathbf{e}_{z}, 
	\\ 
	\partial V_{3}(t) & = \{ (r,\varphi,z) | r \in [0,R(h(t))], \varphi \in [0,2 \pi), z = h(t) \}, 
	& \quad \mathbf{n}_{3} & = \mathbf{e}_{z}. 
	\nonumber
\end{alignat}
Taking into account the local analysis and approximations that are the consequences of \eqref{eq::OrderOfMagnitude}, it turns out that the following results hold: 
\begin{align} 
	\int_{V(t)} \rho \mathbf{v} \, \mathrm{d}V & = 
	\rho \pi h(t) \overline{v}_{h}(t) R(h(t))^{2} \mathbf{e}_{z} + \mathcal{O}(\alpha), 
	\label{eq::GlobalIntegrals-Momentum} \\ 
	\int_{\partial V(t)} \mathbf{t}(\mathbf{n}) \, \mathrm{d}S & = 
	- \pi \left[ R(h(t))^{2} p(h(t)) - R(0)^{2} p(0) \right] \mathbf{e}_{z} 
	- 8 \pi \mu h(t) \overline{v}_{h}(t) \mathbf{e}_{z} + \mathcal{O}(\alpha), 
	\label{eq::GlobalIntegrals-ContactForce} \\ 
	\int_{V(t)} \rho \mathbf{b} \, \mathrm{d}V & = 
	- \rho g \pi h(t) R(h(t))^{2} \mathbf{e}_{z} + \mathcal{O}(\alpha). 
	\label{eq::GlobalIntegrals-BodyForce}
\end{align} 
In particular, equation \eqref{eq::GlobalIntegrals-Momentum} determines the momentum of the fluid contained in $V(t)$. Its derivation relies on the estimate \eqref{eq::EstimateFlux} and the mean velocity of the meniscus \eqref{eq::AverageVelocity}. Equation \eqref{eq::GlobalIntegrals-ContactForce} gives the contact forces over the boundary $\partial V(t)$. It comprises the difference of forces between the lower and the upper base, and effects of viscous stresses on the pipe wall. Crucial roles are played by the estimates \eqref{eq::EstimateVelocity} and \eqref{eq::AverageVelocity}. Finally, \eqref{eq::GlobalIntegrals-BodyForce} provides the body force of the whole fluid volume in $V(t)$. To that end, estimate $\eqref{eq::EstimateRadius}$ is used. Proof of these results is given in Appendix. 

When terms $\mathcal{O}(\alpha)$ are neglected, all expressions in \eqref{eq::GlobalIntegrals-Momentum}-\eqref{eq::GlobalIntegrals-BodyForce} are reduced to their $\mathbf{e}_z$ components. Therefore, consequence of the global momentum balance law \eqref{eq::MomentumGlobal} up to terms of order $\mathcal{O}(\alpha)$ is the following equation: 
\begin{align} \label{eq::PreGeneralisedWashburn}
	\frac{\mathrm{d}}{\mathrm{d}t} \left[ \rho \pi h(t) \overline{v}_{h}(t) R(h(t))^{2} \right] = 
	& - \pi \left[ R(h(t))^{2} p(h(t)) - R(0)^{2} p(0) \right] 
	\\
	& - 8 \pi \mu h(t) \overline{v}_{h}(t) - \rho g \pi h(t) R(h(t))^{2}. 
	\nonumber 
\end{align}
Derivation of the generalized Washburn equation requires a few more steps: introduction of the surface tension as the main driving agent for capillary flow, discarding of still present $\mathcal{O}(\alpha)$ terms, and relation between the mean velocity $\overline{v}_{h}(t)$ and height $h(t)$ of the fluid column. This will be done in the sequel. 

\subsection{The generalized Washburn equation} 

Surface tension coefficient $\gamma$ in narrow pipes may be approximated as $\gamma = F/d$, where $F$ is the magnitude of force due to surface tension and $d$ is length of the line along which the force $F$ acts. In our case, force due to surface tension acts along inner circumference of the pipe, $d = 2 \pi R(h(t))$. It can be shown (see Appendix) that force due to surface tension may be locally decomposed as: 
\begin{equation*}
	\begin{split}
		\mathbf{F} & = F \operatorname{cos} \theta \, \mathbf{e}_z 
		+ F \operatorname{sin} \theta \, \mathbf{e}_r
		+ \mathcal{O}(\alpha) \\
		& = F_{\mathrm{vert}} \, \mathbf{e}_z 
		+ F_{\mathrm{hor}} \, \mathbf{e}_r 
		+ \mathcal{O}(\alpha), 
	\end{split}
\end{equation*}
where $\theta$ is the angle between the tangent to the fluid free surface and axis of the pipe. It is the vertical component of the force, $F_{\mathrm{vert}} = F \operatorname{cos} \theta$, which balances the pressure difference in $\mathbf{e}_{z}$ direction, yielding: 
\begin{equation} \label{eq::SurfaceTension}
	- \pi \left[ R(h(t))^{2} p(h(t)) - R(0)^{2} p(0) \right] 
	= 2 \pi \gamma R(h(t)) \operatorname{cos} \theta. 
\end{equation} 

Rate of change of the momentum hides one extra term of order $O(\alpha)$ which has to be neglected for the sake of consistency: 
\begin{align} \label{eq::MomentumRateApprox}
	\frac{\mathrm{d}}{\mathrm{d}t} \left[ \rho \pi h(t) \overline{v}_{h}(t) R(h(t))^{2} \right] 
	  = & \frac{\mathrm{d}}{\mathrm{d}t} \left[ \rho \pi h(t) \overline{v}_{h}(t) \right] R(h(t))^{2} 
	\nonumber
	\\
	  & + 2 \rho \pi h(t) \overline{v}_{h}(t) \underbrace{R'(h(t))}_{\mathcal{O}(\alpha)} \frac{\mathrm{d}h(t)}{\mathrm{d}t}  
	\\ 
	  = & \frac{\mathrm{d}}{\mathrm{d}t} \left[ \rho \pi h(t) \overline{v}_{h}(t) \right] R(h(t))^{2}
	  + \mathcal{O}(\alpha)
	\nonumber
\end{align}

Finally, momentum balance equation \eqref{eq::PreGeneralisedWashburn} still inherits two unknown functions, $\overline{v}_{h}(t)$ and $h(t)$, unrelated so far. Having in mind that $\overline{v}_{h}(t)$ is the mean velocity of the fluid at the meniscus, it is natural to assume the following relation: 
\begin{equation} \label{eq::VelocityDefinition}
	\overline{v}_{h}(t) = \frac{\mathrm{d} h(t)}{\mathrm{d} t}. 
\end{equation}

Substituting \eqref{eq::SurfaceTension}, \eqref{eq::MomentumRateApprox} and \eqref{eq::VelocityDefinition} into momentum balance equation \eqref{eq::PreGeneralisedWashburn}, we arrive at: 
\begin{align} \label{eq::GeneralisedWashburn}
	\rho \frac{\mathrm{d}}{\mathrm{d}t} 
	  \left[ h(t) \frac{\mathrm{d}h(t)}{\mathrm{d}t} \right] R(h(t))^{2}
	& + 8 \mu h(t) \frac{\mathrm{d} h(t)}{\mathrm{d} t}
	\\ 
	& + \rho g \pi h(t) R(h(t))^{2} 
	= 2 \gamma R(h(t)) \operatorname{cos} \theta, 
	\nonumber
\end{align}
which will be referred to as \emph{the generalized Washburn equation} in the sequel. If the circular cross section of the pipe were constant, with radius equal to the radius of the base, $R(z) = R(0) = R_{0}$, then the generalized Washburn equation \eqref{eq::GeneralisedWashburn} reduces to the standard Washburn equation \eqref{Intro:Washburn}. 

\paragraph{Remark about the initial conditions} The usual initial condition $h(0) = 0$, which is also physically appealing, gives rise to a singularity. It appears in the inertial term and prevents regular solution of the problem. In order to avoid that, we propose $h(0) = c$, $0 < c \ll 1$, like in \cite{rapajic2025134842}, to facilitate numerical computations. Additionally, we suppose that initial velocity equals zero, i.e. $\overline{v}_{h}(0) = \mathrm{d}h(0)/\mathrm{d}t = 0$. 

\subsection{Lubrication theory model and its approximation} 

At first sight, generalized Washburn equation \eqref{eq::GeneralisedWashburn} and lubrication theory model \eqref{Intro:Lubrication} can hardly be compared. On one hand, physical origin of particular terms is quite obvious: both equations contain viscous, gravitational and capillary terms, whereas lubrication theory model lacks inertial term. On the other hand, common terms have quite different structure. Therefore, a question may be raised why these two model, based upon similar assumptions, differ so much in their structure? It will be revealed in the sequel that difference between them is not so big. It will be even quantified---either analytically, or numerically, we shall end up with the difference of order $\mathcal{O}(\alpha)$. 

The first step in comparison will be of analytical kind. It is based upon fundamental assumption \eqref{eq::OrderOfMagnitude}, $|R'(z)| = \mathcal{O}(\alpha) \ll 1$, and the estimate \eqref{eq::EstimateRadius}. In particular, lubrication theory model requires two estimates. First, from \eqref{eq::EstimateRadius} it follows: 
\begin{equation*}
	R(h(t))^{3} \int_{0}^{h(t)} R(z)^{-4} \, \mathrm{d}z 
	= h(t) R(h(t))^{-1} + \mathcal{O}(\alpha). 
\end{equation*}
Second, since $R'(z) = \tan \alpha$, using \eqref{eq::OrderOfMagnitude} one obtains: 
\begin{equation}
	\mathrm{cos} \left( \theta + \mathrm{arctan} \frac{\mathrm{d}R(h(t))}{\mathrm{d}h} \right) 
	= \mathrm{cos} (\theta + \alpha) = \operatorname{cos} \theta + \mathcal{O}(\alpha).
	\label{eq::CosApprox}
\end{equation}
Substituting last two estimates into \eqref{Intro:Lubrication}, and neglecting $\mathcal{O}(\alpha)$ terms, the approximate lubrication theory model is obtained: 
\begin{equation} \label{eq::Lubrication-Approx}
	8 \mu R(h(t))^{-1} h(t) \frac{\mathrm{d} h(t)}{\mathrm{d} t} 
	+ \rho g h(t) R(h(t)) = 2 \gamma \operatorname{cos} \theta. 
\end{equation}
Assuming that pipe is open at the top, i.e. $R(h(t)) > 0$, simple inspection reveals that approximate lubrication theory model \eqref{eq::Lubrication-Approx} becomes the special case of the generalized Washburn equation \eqref{eq::GeneralisedWashburn} in which the inertial term is neglected. 

\subsection{Equilibrium height}

This part of the analysis will be concluded with a simple estimate of the equilibrium height. Generalized Washburn's equation \eqref{eq::GeneralisedWashburn}, and the approximate lubrication theory model \eqref{eq::Lubrication-Approx}, imply the same equation for equilibrium height $\overline{h}$. 
\begin{equation}
	\rho g \pi \overline{h} R(\overline{h}) = 2 \gamma \cos \theta. 
\end{equation} 
On the other hand, equation \eqref{Intro:Washburn} for the pipe with constant circular cross section with radius $R_{0} = R(0)$ yields the well known equilibrium height $h_{\mathrm{e}}$, usually referred to as the Jurin height: 
\begin{equation} \label{eq::HeightJurin}
	h_{\mathrm{e}} = \frac{2 \gamma \operatorname{cos} \theta}{\rho g R_{0}}. 
\end{equation}
Putting these two results together, the following relation is obtained: 
\begin{equation*}
	\overline{h} R(\overline{h}) 
	= \frac{2 \gamma \operatorname{cos} \theta}{\rho g} 
	= h_{\mathrm{e}} R_{0}. 
\end{equation*}
If the analysis is restricted to converging pipes, $R(z) < R(0) = R_{0}$, a simple estimate is obtained: 
\begin{equation} \label{eq::h-estimate}
	\frac{\overline{h}}{h_{\mathrm{e}}} = \frac{R_{0}}{R(\overline{h})} > 1.
\end{equation}
At the same, volume of the fluid that fills the converging pipe in equilibrium is: 
\begin{equation*}
	V(\overline{h}) = \int_{0}^{\overline{h}} \int_{0}^{R(z)} \int_{0}^{2\pi} 
	  r \, \mathrm{d}r \, \mathrm{d}\varphi \, \mathrm{d}z 
	  = \overline{h} R(\overline{h})^{2} \pi + \mathcal{O}(\alpha),
\end{equation*}
whereas the volume of the fluid that fills a constant radius pipe is $V_{0} = h_{e} R_{0}^{2} \pi$. Neglecting $\mathcal{O}(\alpha)$, these results in conjunction with \eqref{eq::h-estimate} lead to: 
\begin{equation}
	\frac{V(\overline{h})}{V_{0}} 
	  = \frac{\overline{h} R(\overline{h})^{2} \pi}{h_{e} R_{0}^{2} \pi} 
	  = \frac{R(\overline{h})}{R_{0}} < 1. 
\end{equation}
It may be concluded that equilibrium height $\overline{h}$ in a converging pipe with a base radius $R_{0}$ is greater than the Jurin height in a cylindrical pipe of radius $R_{0}$. However, equilibrium volume of the fluid in a converging pipe is smaller than the volume in a cylindrical one. 

\subsection{Dimensionless models} 

\paragraph{Dimensionless variables} To gain deeper insight into the structure of the models, and to facilitate their numerical solution, they will be put into dimensionless form. For a reference length scale it will be used Jurin height \eqref{eq::HeightJurin}, that is equilibrium height in the cylindrical pipe of radius $R_{0}$. Reference time scale $\tau$ is determined during the scaling process such that coefficient of the viscous term equals 1. Therefore, we have: 
\begin{equation} \label{eq::DLessVariables}
	H = \frac{h}{h_{\mathrm{e}}}, \; Z = \frac{z}{h_{\mathrm{e}}}, \; 
	T = \frac{t}{\tau}, \; \tau = \frac{8 \mu h_\mathrm{e}}{\rho g R_0^2}. 
\end{equation}
Function $R(z)$, that describes variable radius of the circular cross section, is scaled using auxiliary dimensionless function $\overline{R}(z)$: $R(z) = R_{0} \overline{R}(z)$, $\overline{R}(0) = 1$. In such a way we may introduce dimensionless radius function $\widehat{R}(Z)$:
\begin{equation} \label{eq::DLessRadius}
	\widehat{R}(Z) = \overline{R}(h_{\mathrm{e}} Z) = \frac{R(h_{\mathrm{e}} Z)}{R_{0}}. 
\end{equation}
Note that fundamental assumption \eqref{eq::OrderOfMagnitude} of our analysis is preserved for dimensionless radius function:
\begin{equation*}
	\left| \widehat{R}'(Z) \right| = |\tan \alpha| = \mathcal{O}(\alpha) \ll 1. 
\end{equation*}

After scaling there will remain only one (dimensionless) parameter in the model introduced in \cite{Switala}: 
\begin{equation} \label{eq::Omega}
	\omega = \frac{h_{\mathrm{e}}}{g \tau^{2}} = \frac{\rho^2 R_0^4 g}{64 \mu^2 h_\textrm{e}}. 
\end{equation}
In \cite{Bothe} it was given a nice interpretation of $\omega$ in terms of another two dimensionless numbers, the Ohnesorge number $\mathrm{Oh}$ and the Bond number $\mathrm{Bo}$: 
\begin{align}\label{eq::omega_Bo-Oh}
	\omega & = \frac{1}{128 \operatorname{cos}\theta} \left( \frac{\mathrm{Bo}}{\mathrm{Oh}} \right)^{2},    
	\\ 
	\mathrm{Oh} & = \frac{\mu}{\sqrt{R_{0}\rho\gamma}} = \frac{\sqrt{\mathrm{We}}}{\mathrm{Re}}, 
	\quad \mathrm{Bo} = \frac{\rho g R_{0}^{2}}{\gamma}. 
	\nonumber
\end{align}
The Ohnesorge number relates viscosity to inertia and surface tension, and can be expressed as a ratio of square root of Weber number to Reynolds number. The Bond (E\"{o}tv\"{o}s) number relates gravitational forces to surface tension. In \cite{rapajic2025134842}, it is given a detailed study of the reduced models obtained by means of asymptotic analysis (either for $\omega \to 0$, or $\omega \to \infty$) in the case of a constant radius cylindrical pipe. 

\paragraph{Generalized Washburn's equation} Using dimensionless variables, the generalized Washburn equation \eqref{eq::GeneralisedWashburn} can be transformed into dimensionless form: 
\begin{equation} \label{eq::DLessGeneralisedWashburn}
	\omega \, \frac{\textrm{d}}{\textrm{d}T} \left[ H(T) 
	\frac{\textrm{d}H(T)}{\textrm{d}T} \right] \widehat{R}(H(T))^{2} 
	+ H(T) \frac{\textrm{d}H(T)}{\textrm{d}T} 
	+  H(T) \widehat{R}(H(T))^{2} = \widehat{R}(H(T)).
\end{equation} 

\paragraph{Lubrication theory model} In the same way as for the generalized Washburn equation, lubrication theory model \eqref{Intro:Lubrication} transformed into dimensionless form reads: 
\begin{equation} \label{eq::DLessLubrication}
	\left( \widehat{R}(H(T))^{3} \int_{0}^{H(T)} \widehat{R}(Z)^{-4} \textrm{d}Z \right) 
	\frac{\textrm{d}H(T)}{\textrm{d}T} + H(T) \widehat{R}(H(T)) = 1, 
\end{equation}
whereas its approximate counterpart \eqref{eq::Lubrication-Approx} has the form:
\begin{equation} \label{eq::DLessLubrication-Approx}
	\widehat{R}(H(T))^{-1} H(T) \frac{\textrm{d}H(T)}{\textrm{d}T} 
	+ H(T) \widehat{R}(H(T)) = 1. 
\end{equation}
Again, \eqref{eq::DLessLubrication-Approx} may be derived from \eqref{eq::DLessLubrication} through order of magnitude analysis. 

At this point, comparison of dimensionless models \eqref{eq::DLessGeneralisedWashburn} and \eqref{eq::DLessLubrication-Approx} reveals that lubrication theory model in approximate form may be recovered from the generalized Washburn equation for $\omega \to 0$. 


\section{Comparison of the numerical results} \label{sec:ComparisonNum} 


In this section, we perform numerical simulations where the radius of the pipe varies with the $z$-coordinate according to the following expression:
\begin{equation*}
	R(z) = R_0 (1 + z/L)^{-0.4}.
\end{equation*}
In addition, we choose the length of the pipe to be 20 times the equilibrium height. This ensures that the fluid has enough length to reach equilibrium without overflowing the pipe.

It is a remarkable result of the capillary rise analysis, both experimental \cite{Quere} and theoretical \cite{Zhmud}, that the equilibrium height may be approched monotonically or oscillatory in the pipe of constant radius. This turned out to depend on the material properties. The dynamics of the capillary rise was explained \cite{Switala} by the stability properties of the stationary point corresponding to the equilibrium height. Recently, it was proved \cite{rapajic2025134842} that under physically reasonable assumptions every solution of Washburn's equation, which starts from a small non-negative height and from rest, converges to the equilibrium height in this way. Naturally, there comes the question of asymptotic behavior in the pipe with variable radius. 
\begin{figure}[h!]
	\includegraphics[scale=0.7]{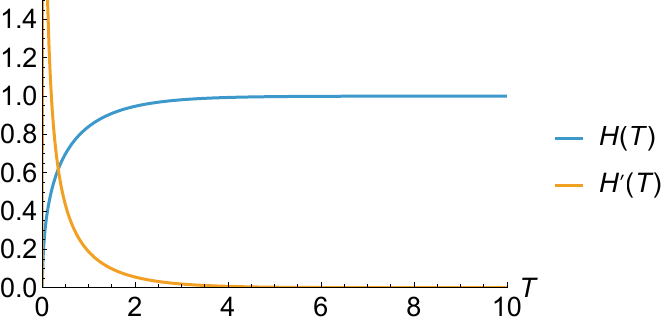}
	\includegraphics[scale=0.7]{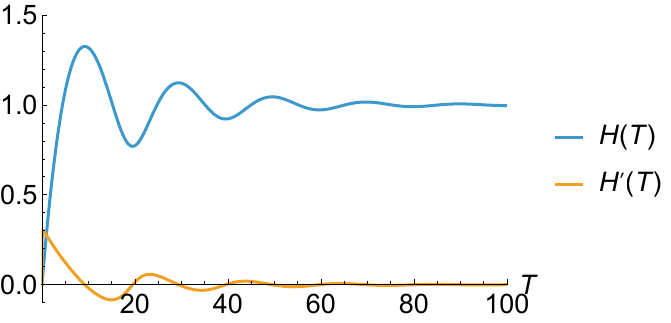}
	\caption{Solution $H(T)$ of the generalized Washburn equation \eqref{eq::DLessGeneralisedWashburn} for different values of critical parameter; $L = 20h_\textrm{e}, R_0 = 1$. Left: $\omega = 0.01$; right: $\omega = 10$.}
	\label{fig::monotoneVSoscillatory}
\end{figure}

Figure \ref{fig::monotoneVSoscillatory} shows that, in the pipe with a variable radius, the equilibrium height can be reached either monotonically or oscillatory, as in the constant radius case. Specifically, for small values of the parameter $\omega$, the equilibrium height is approached monotonically, while for larger values of $\omega$ the equilibrium height is reached in an oscillatory manner. In other words, our hypothesis is that the equilibrium point behaves as a stable node or a stable focus. 

\begin{figure}[h!]
	\begin{center}
		\includegraphics[scale=0.5]{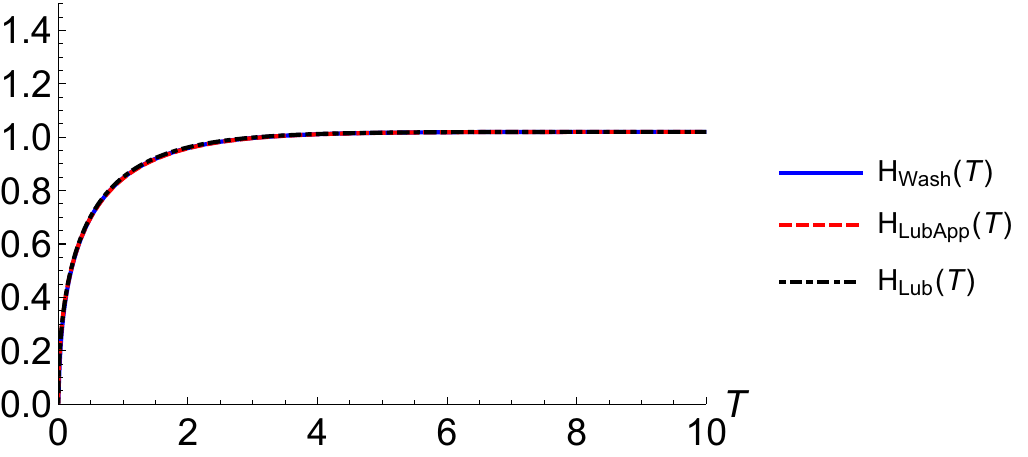}
		\includegraphics[scale=0.5]{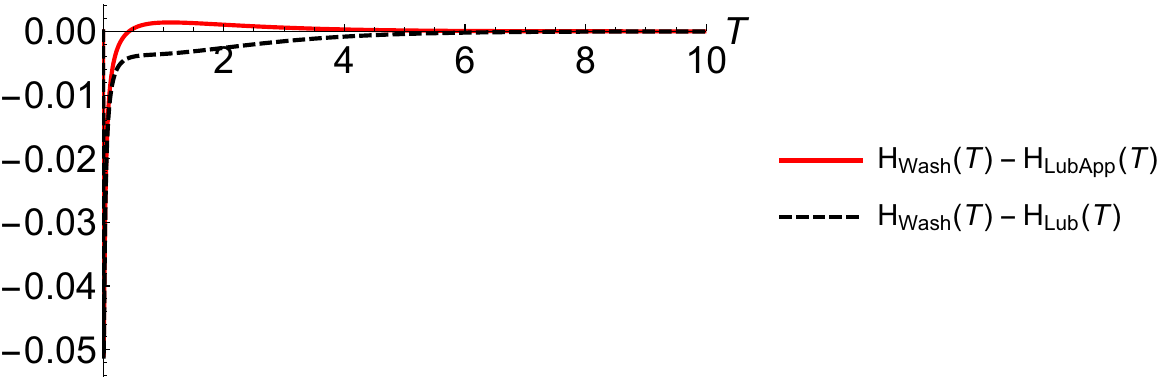}
	\end{center}
	\caption{Comparison of the solution $H_{\mathrm{Wash}}(T)$ of the generalized Washburn equation \eqref{eq::DLessGeneralisedWashburn} to the solution $H_{\mathrm{LubApp}}(T)$ \eqref{eq::DLessLubrication-Approx} and $H_{\mathrm{Lub}}(T)$ \eqref{eq::DLessLubrication}, $L = 20 h_\textrm{e}$, $\omega = 0.01$.}
	\label{fig::WashVSLub-Small}
\end{figure}
Numerical solution of the generalized Washburn's equation \eqref{eq::DLessGeneralisedWashburn} is compared to lubrication theory model \eqref{eq::DLessLubrication} and its approximation \eqref{eq::DLessLubrication-Approx}. For the small values of $\omega$ (Figure \ref{fig::WashVSLub-Small}), the difference between solutions is of the order $\mathcal{O}(\alpha)$. It suggests that our model is sufficiently accurate and thus suitable for practical applications. The advantage of our model is that it allows a more transparent interpretation of each term, making it significantly more suitable for theoretical analysis, such as establishing existence and uniqueness of solutions, similarly to what was done in \cite{rapajic2025134842}.

\begin{figure}[h!]
	\begin{center}
		\includegraphics[scale=0.5]{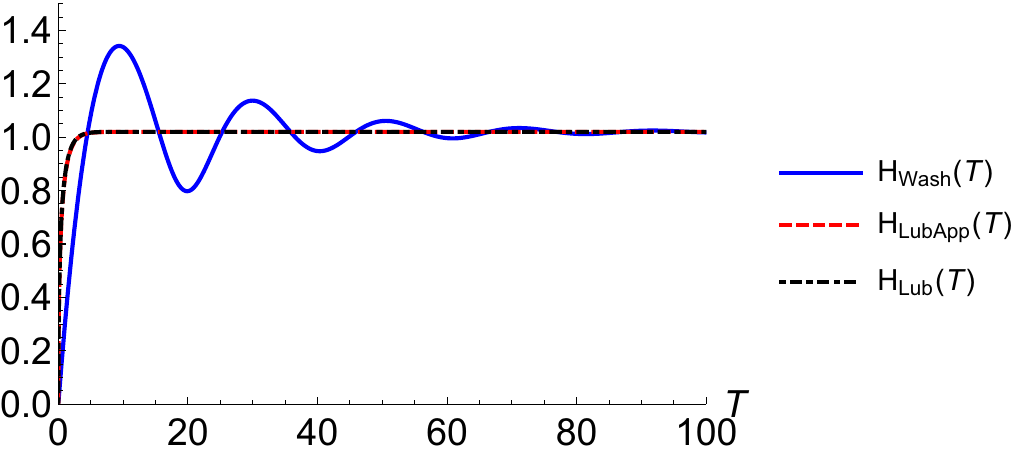}\\
		\includegraphics[scale=0.5]{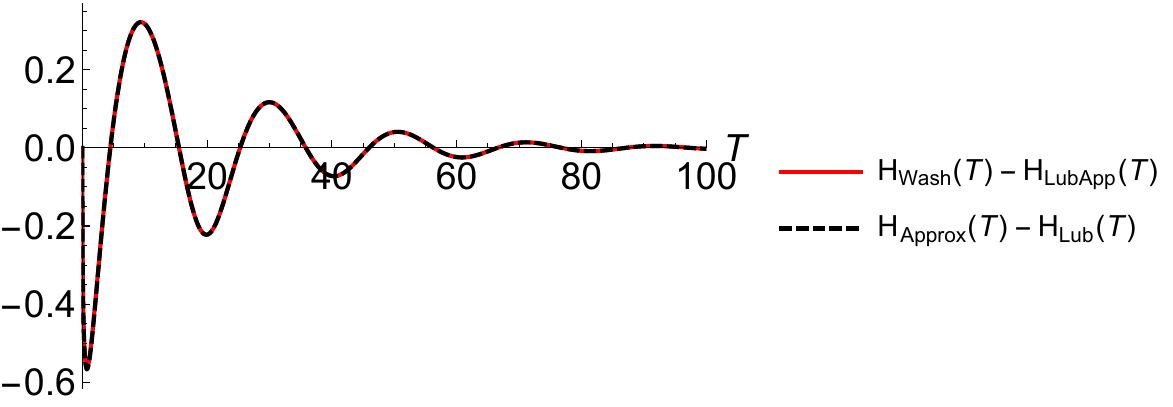}
	\end{center}
	\caption{Comparison of the solution $H_{\mathrm{Wash}}(T)$ of the generalized Washburn equation \eqref{eq::DLessGeneralisedWashburn} to the solution $H_{\mathrm{LubApp}}(T)$ \eqref{eq::DLessLubrication-Approx} and $H_{\mathrm{Lub}}(T)$ \eqref{eq::DLessLubrication}, $L = 20 h_\textrm{e}$, $\omega = 10$.}
	\label{fig::WashVSLub-Large}
\end{figure}
In Figure \ref{fig::WashVSLub-Large}, we show that for the large values of $\omega$ the complete model \eqref{eq::DLessGeneralisedWashburn} becomes essential---lubrication theory contains no dependence on $\omega$ and therefore cannot capture the oscillatory behaviour. Although this follows implicitly from the formulation itself, we present the numerical confirmation to emphasize the necessity for generalization of the Washburn equation. 


\section{Conclusion}


In this study, derivation of the capillary rise equation is presented for a pipe with variable cross-section. The main result is the new model, generalized Washburn equation \eqref{eq::GeneralisedWashburn}. It generalizes an ODE model---Washburn's equation \eqref{Intro:Washburn}---and it is compared with a model \eqref{Intro:Lubrication} based upon lubrication theory. The peculiarity of the approach adopted in this paper is that it is based upon first principles, i.e. application of the mass and momentum balance laws in local form (Navier-Stokes equations) and global form (continuity and momentum equation). Along with a precise list of assumptions and careful application of averaging procedures (global analysis) and order of magnitude estimates (asymptotic analysis), this allowed for a better control over the terms in the resulting equation. 

The distinguishing feature of our model is the presence of an inertial term, which is excluded from the lubrication theory model by assumption. This allowed for a comparison between the generalized Washburn equation and the lubrication theory model. Namely, the lubrication theory model can be recovered when the parameter $\omega$ vanishes in the dimensionless generalized model. 

Numerical simulations confirmed that our generalized Washburn's equation aligns well with both its approximation and the lubrication theory model, providing a form of verification. By the analysis of numerical solutions, the following important features were revealed. 
\begin{itemize}
	\item[(i)] The generalized Washburn equation preserves both monotonic and oscillatory approach to equilibrium height in a slowly converging pipe, depending on the value of dimensionless parameter $\omega$. 
	
	\item[(ii)] The difference between the solutions of generalized Washburn's equation \eqref{eq::DLessGeneralisedWashburn} and the solutions of exact and approximate lubrication theory equations is of the order of magnitude of an inclination angle of the pipe, which is assumed to be small. 
	
	\item[(iii)] During the motion, difference between the solutions of the generalized Washburn equation and the lubrication theory model is of the order of magnitude of a small parameter $\omega$, in the case when equilibrium height is reached monotonically.
\end{itemize}

This work opens up many possibilities for further analysis. In particular, the generalized Washburn equation may be analyzed for the existence and uniqueness of solutions, stability of equilibrium, and comparison with experimental results. In the long run, this study may be a starting point for capillary flow analysis without restrictions on the shape of the pipe, or the monotonicity of radius variation. 


\appendix 


\section{Local decomposition of the surface tension}

We locally decompose surface tension using the approximation \eqref{eq::CosApprox}.
\begin{equation*}
	\mathbf{F} = F \operatorname{cos} \left( \theta + \alpha \right)  \mathbf{e}_z + F \operatorname{sin}\left( \theta + \alpha \right) \mathbf{e}_r
\end{equation*}
This expression is further transformed by applying the addition formulas for sine and cosine
\begin{equation*}
	\begin{split}
		\mathbf{F} & = F \left[ \operatorname{cos} \theta \operatorname{cos} \alpha - \operatorname{sin} \theta \operatorname{sin} \alpha \right] \mathbf{e}_z + 
		F \left[ \operatorname{sin} \theta \operatorname{cos} \alpha + \operatorname{cos} \theta \operatorname{cos} \alpha  \right] \mathbf{e}_r \\
		& = F \left[ \operatorname{cos} \theta + \mathcal{O}(\alpha) \right]  \mathbf{e}_z + F \left[ \operatorname{sin} \theta + \mathcal{O}(\alpha)  \right]  \mathbf{e}_r,
	\end{split}    
\end{equation*}
In other words,
\begin{equation*}
	\mathbf{F} = F \operatorname{cos} \theta \mathbf{e}_z + F  \operatorname{sin} \theta \mathbf{e}_r + \mathcal{O}(\alpha) =: F_{\mathrm{vert}} \mathbf{e}_z +  F_{\mathrm{hor}} \mathbf{e}_r + \mathcal{O}(\alpha)
\end{equation*}

\section{Global momentum} 

Global momentum of the fluid is computed using the velocity field \eqref{eq::VelocityFieldApprox} as follows: 
\begin{equation*}
\begin{split}
	\int_{V(t)} \rho \mathbf{v} \, \mathrm{d}V 
	  & = \left[ \int_{V(t)} \rho v_{z}(r,z,t) \, \mathrm{d}V \right] \mathbf{e}_{z} 
	  + \mathcal{O}(\alpha) 
	\\ 
	& = \left[ \int_{0}^{h(t)} \int_{0}^{R(z)} \int_{0}^{2\pi} \rho v(z,t) 
	  \left( 1 - \frac{r^{2}}{R(z)^{2}} \right) r \mathrm{d}r \mathrm{d}\varphi \mathrm{d}z \right] 
	  \mathbf{e}_{z} + \mathcal{O}(\alpha) 
	\\ 
	& = \frac{1}{2} \rho \pi \left[ \int_{0}^{h(t)} v(z,t) R(z)^{2} \mathrm{d}z \right] 
	  \mathbf{e}_{z} + \mathcal{O}(\alpha) 
	\\ 
	& = \frac{1}{2} \rho \pi h(t) v(h(t),t) R(h(t))^{2} \mathbf{e}_{z} + \mathcal{O}(\alpha).  
\end{split}
\end{equation*}
In the last step, we took into account the global flux estimate \eqref{eq::EstimateFlux}. The final form of global momentum is obtained when the modulating factor $v(h(t),t)$ is substituted by the mean velocity of the meniscus \eqref{eq::AverageVelocity}: 
\begin{equation*}
	\int_{V(t)} \rho \mathbf{v} \, \mathrm{d}V 
	  = \rho \pi h(t) \overline{v}_{h}(t) R(h(t))^{2} \mathbf{e}_{z} + \mathcal{O}(\alpha).
\end{equation*}

\section{Contact force and body force} 

The total contact force $\mathbf{F}_s$ is determined by integrating traction over the pipe boundary, which is divided into three regions \eqref{eq::Boundary}. 

\subsection{Contact force integrals}

To compute the contact force, we use the following expression for traction \cite{acheson1990elementary}, valid for incompressible viscous fluids
\begin{equation}
	\mathbf{t}(\mathbf{n}) = - p \mathbf{n} + \mu \left[ 2 (\mathbf{n} \cdot \nabla) \mathbf{v} + \mathbf{n} \times (\nabla \times \mathbf{v}) \right]. 
	\label{eq::StressAcheson}
\end{equation} 
Surface integral over the lower base:
\begin{equation*}
	\int_{\partial V_1} \mathbf{t}_1 \mathrm{d}S_1 = \int_0^{R_0} \int_0^{2 \pi} \mathbf{t}_1 r \mathrm{d}r \mathrm{d} \varphi = p(0)\mathbf{e}_z \int_0^{R_0} \int_0^{2 \pi} r \mathrm{d}r \mathrm{d}\varphi = R_0^2 \pi p(0) \mathbf{e}_z.
\end{equation*}
Surface integral over the upper base:
\begin{equation*}
	\int_{\partial V_3} \mathbf{t}_3 \mathrm{d}S_3 = \int_0^{R(h(t))} \int_0^{2\pi} -p(h(t))\mathbf{e}_z r \mathrm{d}r \mathrm{d}\varphi = -R(h(t))^{2} \pi p(h(t))\mathbf{e}_z.
\end{equation*}
Adding up these two integrals we obtain vertical force which will be balanced with vertical component of surface tension:
\begin{equation*}
	\int_{\partial V_1} \mathbf{t}_1 \mathrm{d}S_1 + \int_{\partial V_3} \mathbf{t}_3 \mathrm{d}S_3 
	= \left( R_0^2 \pi p(0) - R(h(t))^{2} \pi p(h(t)) \right) \mathbf{e}_z =: F_{\text{vert}}\mathbf{e}_z.
\end{equation*}
We assume that
\begin{equation*}
	F_{\text{vert}} = 2 R(h(t)) \pi \gamma \operatorname{cos} \theta.
\end{equation*}
Thus,
\begin{equation*}
	\int_{\partial V_1} \mathbf{t}_1 \mathrm{d}S_1 
	+ \int_{\partial V_3} \mathbf{t}_3 \mathrm{d}S_3 
	= 2 R(h(t)) \pi \gamma \operatorname{cos} \theta.
\end{equation*}

\subsubsection{Contact force over $\partial V_2$}

To compute this part of the contact force, approximate form of the velocity field \eqref{eq::VelocityFieldApprox}, with Poiseuille flow profile \eqref{eq::PoiseuilleFlow}, will be used
\begin{equation*}
	\mathbf{v} = v(z,t) \left( 1 - \frac{r^{2}}{R(z)^{2}} \right) \mathbf{e}_z + \mathcal{O}(\alpha). 
\end{equation*}
Also, the unit normal vector will be approximated as 
\begin{equation*}
	\mathbf{n}_2 = \operatorname{cos} \alpha \mathbf{e}_r + \operatorname{sin} \alpha \mathbf{e}_z
	= \mathbf{e}_r + \mathcal{O}(\alpha). 
\end{equation*}
Taking into account these approximations and using them in \eqref{eq::StressAcheson}, leads to 
\begin{equation*}
	\begin{split}
		\int_{\partial V_{2}(t)} \mathbf{t}_2 \mathrm{d}S_2 
		& = \int_{\partial V_{2}(t)} \left[ -p(z) \mathbf{e}_r - 2 \mu \frac{v(z,t)}{R(z)} \mathbf{e}_z \right] R(z) \mathrm{d}\varphi \mathrm{d}z + \mathcal{O}(\alpha) \\
		& = - \int_0^{h(t)} p(z) R(z) \mathrm{d}z \underbrace{\int_0^{2\pi} \mathbf{e}_r \mathrm{d}\varphi}_{=0} - 2 \mu \left[ \int_0^{h(t)} v(z,t) \mathrm{d}z 
		\int_0^{2\pi} \mathrm{d}\varphi \right] \mathbf{e}_z + \mathcal{O}(\alpha) \\ 
		& = - 4 \pi \mu h(t) v(h(t),t) \mathbf{e}_{z} + \mathcal{O}(\alpha) \\
		& = - 8 \pi \mu h(t) \overline{v}_{h}(t) \mathbf{e}_z + \mathcal{O}(\alpha),    
	\end{split}
\end{equation*}
where we used the mean velocity \eqref{eq::AverageVelocity}, $\overline{v}_{h}(t) = v(h(t),t)/2$. 

\subsection{Body force integral}

Body force is only due to gravity, $\mathbf{b} = \mathbf{g} = - g \mathbf{e}_z$. with this, total body force acting on volume $V(t)$ reads
\begin{equation*}
	\begin{split}
		\int_{V(t)} \rho \mathbf{b} \mathrm{d}V 
		& = - \rho \left( \int_0^{h(t)} \int_0^{R(z)} \int_0^{2 \pi} g r \mathrm{d}r\mathrm{d}\varphi \mathrm{d}z \right) \mathbf{e}_z \\ 
		& = -\rho \left[ \int_0^{h(t)} \left( \int_0^{R(z)} g r \mathrm{d}r \int_0^{2 \pi} \mathrm{d}\varphi \right) \mathrm{d}z \right] \mathbf{e}_z \\
		& = - \rho g \pi \left[ \int_0^{h(t)} R(z)^2 \mathrm{d}z \right] \mathbf{e}_z. 
	\end{split}
\end{equation*}
Using the estimate \eqref{eq::EstimateRadius}, the following expression for the body force is obtained: 
\begin{equation*}
	\int_{V(t)} \rho \mathbf{b} \mathrm{d}V = - \rho g \pi h(t) R(h(t))^{2} \mathbf{e}_{z} 
	  + \mathcal{O}(\alpha). 
\end{equation*}

\bibliographystyle{elsarticle-num} 
\bibliography{references.bib}






\end{document}